\documentclass[conference]{IEEEtran}
\IEEEoverridecommandlockouts
\usepackage{graphicx}
\usepackage{svg}
\usepackage{tikz}
\usepackage{comment}
\usepackage{algorithm}
\usepackage{algpseudocode}
\usepackage{caption}
\usepackage{subcaption}
\usepackage{siunitx}
\usepackage{cite}
\usepackage{amsmath,amssymb,amsfonts}
\usepackage{graphicx}
\usepackage{textcomp}
\usepackage{xcolor}
\def\BibTeX{{\rm B\kern-.05em{\sc i\kern-.025em b}\kern-.08em
    T\kern-.1667em\lower.7ex\hbox{E}\kern-.125emX}}
\algnewcommand\algorithmicforeach{\textbf{for each}}
\algdef{S}[FOR]{ForEach}[1]{\algorithmicforeach\ #1\ \algorithmicdo}

\newcommand{\daniele}[1]{\textcolor{blue}{Daniele: #1}}
\newcommand{\luca}[1]{\textcolor{orange}{Luca: #1}}
\newcommand{\francesca}[1]{\textcolor{red}{Francesca: #1}}

\begin{document}
% \renewcommand\thelinenumber{\color[rgb]{0.2,0.5,0.8}\normalfont\sffamily\scriptsize\arabic{linenumber}\color[rgb]{0,0,0}}
% \renewcommand\makeLineNumber {\hss\thelinenumber\ \hspace{6mm} \rlap{\hskip\textwidth\ \hspace{6.5mm}\thelinenumber}}
% \linenumbers
%\pagestyle{headings}
%\mainmatter

\title{Exploiting Liver CT scans in Colorectal Carcinoma genomics mutation classification} % Replace with your title

% INITIAL SUBMISSION 
%\begin{comment}
%\titlerunning{} 
%\authorrunning{} 
%\author{}
\author{\IEEEauthorblockN{Daniele Perlo}
\IEEEauthorblockA{\textit{Luxembourg Institute of Health}, \\
\textit{Fondazione Ricerca Molinette Onlus} \\
daniele.perlo@lih.lu}
\and
\IEEEauthorblockN{Luca Berton}
\IEEEauthorblockA{\textit{Fondazione Ricerca Molinette Onlus} \\
Turin, Italy \\
luca.berton@edu.unito.it}
\and
\IEEEauthorblockN{Alessia Delpiano}
\IEEEauthorblockA{\textit{Fondazione Ricerca Molinette Onlus } \\
Turin, Italy \\
alessia.delpiano3@gmail.com}
\and
\IEEEauthorblockN{Francesca Menchini}
\IEEEauthorblockA{\textit{Dept. of Radiology} \\
\textit{University of Turin}\\
Turin, Italy \\
francesca.menchini@edu.unito.it}
\and
\IEEEauthorblockN{Stefano Tibaldi}
\IEEEauthorblockA{\textit{Dept. of Radiology} \\
\textit{University of Turin}\\
Turin, Italy \\
tibaldistefano1@gmail.com}
\and
\IEEEauthorblockN{Marco Grosso}
\IEEEauthorblockA{\textit{Città della Salute di Torino} \\
Turin, Italy \\
marco.grosso@unito.it}
\and
\IEEEauthorblockN{Paolo Fonio}
\IEEEauthorblockA{\textit{Dept. of Radiology} \\
\textit{University of Turin}\\
Turin, Italy \\
 paolo.fonio@unito.it}
}
%\institute{}
%\end{comment}
%******************

% CAMERA READY SUBMISSION
\begin{comment}
\titlerunning{Exploiting Liver CT scans in Colorectal Carcinoma genomics mutation classification}
% If the paper title is too long for the running head, you can set
% an abbreviated paper title here
%
\author{ Daniele Perlo\inst{1}\orcidID{0000-0001-6879-8475} \and
Luca Berton\inst{1} \and
Alessia Delpiano\inst{1} \and
Francesca Menchini\inst{2} \and
Stefano Tibaldi\inst{2}\orcidID{0000-0003-0326-1399} \and
Marco Grosso\inst{3}\orcidID{0000-0002-7083-09295}
Paolo Fonio\inst{2}\orcidID{0000-0002-1106-2990}}
%
\authorrunning{D. Perlo et al.}
% First names are abbreviated in the running head.
% If there are more than two authors, 'et al.' is used.
%
\institute{Fondazione Ricerca Molinette Onlus \and University of Turin \and Citt\`a della Salute e della Scienza di Torino}
%\institute{Princeton University, Princeton NJ 08544, USA \and
%Springer Heidelberg, Tiergartenstr. 17, 69121 Heidelberg, Germany
%\email{lncs@springer.com}\\
%\url{http://www.springer.com/gp/computer-science/lncs} \and
%ABC Institute, Rupert-Karls-University Heidelberg, Heidelberg, Germany\\
%\email{\{abc,lncs\}@uni-heidelberg.de}}

\end{comment}
%******************
\maketitle

\begin{abstract}
%The abstract should summarize the contents of the paper. LNCS guidelines
%indicate it should be at least 70 and at most 150 words. It should be set in 9-point
%font size and should be inset 1.0~cm from the right and left margins.

The liver is the most involved organ by distant metastasis in colon-rectal cancer (CRC) patients and it comes necessary to be aware of the mutational status of the lesions to correctly design the best individual treatment. So far, efforts have been made in order to develop non-invasive and real-time methods that permit the analysis of the whole tumor, using new artificial intelligence tools to analyze the tumor's image obtained by Computed Tomography (CT) scan. In order to address the current medical workflow, that is biopsy analysis-based, we propose the first DeepLearning-based exploration, to our knowledge, of such classification approach from the patient medical imaging. We propose \textit{i)} a solid pipeline for managing undersized datasets of available CT scans and \textit{ii)} a baseline study for genomics mutation diagnosis support for preemptive patient follow-up. Our method is able to identify CRC RAS mutation family from CT images with 0.73 F1 score. 

%\keywords{We would like to encourage you to list your keywords within the abstract section}
\end{abstract}

\section{Introduction}

Colon-rectal cancer is the third most diagnosed cancer and the second most deadly cancer world-wide~\cite{RAD7} and it generally emerges from glandular and epithelial cells of the large intestine, after the acquisition of genetic or epigenetic mutations~\cite{RAD2.1}. The replicative advantage acquired by these cells leads to the creation of adenoma, which can evolve in carcinoma and give distant metastasis~\cite{Vogelstein_1988}. 
The diagnostic process usually starts from a positive result of the screening test or from the development of typical symptoms; the diagnosis is usually confirmed by a total colonoscopy~\cite{RAD3.1} and followed by the magnetic resonance imaging (MRI) to evaluate the tumor’s local extension and its relationship with the surrounding structures\cite{Sahani_2014}.
Hence, according to the latest guidelines by European Society of Medical Oncology (ESMO), as a preoperative assessment it is highly recommended the execution of a thoracic, abdominal and pelvic CT scan with intravenous contrast~\cite{RAD3.1,Kekelidze_2013}: it is estimated that 20\% of the new diagnosis of colon-rectal cancer already have synchronous metastasis~\cite{RAD3.1}, and almost 60\% will develop them, with the liver being the main affected organ~\cite{RAD4.1}. 
CT is chosen over ultrasounds because of its better performance in detecting liver lesions with 84\% sensitivity and 96\% specificity~\cite{Kekelidze_2013}, and MRI examination can be considered just for ambiguous liver lesions, smaller than 1 cm in size~\cite{RAD3.1,Kekelidze_2013}.

We refer to oligometastatic disease when the distant lesions are localized in two or three sites (that includes liver, lungs, nodes and ovary) with usually up to five lesions in total, even if sometimes they can be even an higher number~\cite{RAD9}. 
As already said, the liver is the most involved organ and 17\% of the synchronous metastases are localized there~\cite{RAD3.1}. In these cases, local treatments are still an option: liver resection is usually the standard, with ablative treatments as alternative solution~\cite{RAD9}.
Research efforts are now focused on demonstrating the equal efficacy of these different techniques, assuming the future introduction of ablative treatments in resectable metastases management, not only as a second option compared to surgery~\cite{Meijerink_2019}.
The assessment of patients with oligometastatic disease must be carried out by a multidisciplinary team, in order to identify the best therapeutic strategy and choose eventual addition of systemic therapy to the local one~\cite{RAD9,Van_Cutsem_2014}. 
A recent study has demonstrated the increasing progression-free survival and overall survival using a combination of local treatment and systemic therapy in patients with unresectable metastasis~\cite{Ruers_2017}.
Before the beginning of the systemic therapy, it is necessary to perform a histological evaluation of the primary tumor or metastasis, in case they have an atypical imaging aspect or if they appear three years after the colon-rectal cancer diagnosis~\cite{Van_Cutsem_2014}. 
In the individual treatment definition, the mutational status of the primary and the secondary localizations has a crucial role as well, having diagnostic, prognostic and predicting value~\cite{Van_Cutsem_2014,RAD10}. 
In particular the mutations involving the epidermal growth factor receptor (EGFR) signaling network are the most investigated ones, due to their importance in the treatment response, and among these the rat sarcoma viral onco-gene (RAS) mutation has a special relevance.
Indeed, RAS is present in almost 40\% of the patients with colon-rectal cancer\cite{RAD11} and the tumors that present this mutation will unlikely respond to the EGFR antibodies therapy~\cite{RAD9,RAD10,RAD12}. Moreover, mutant RAS tumor seems to be more aggressive, with a higher number of positive resection margins after hepatic surgery~\cite{RAD13} and an increased risk of progression after percutaneous treatment with a shorter local tumor progression (LTP)-free survival~\cite{RAD12,RAD14}. 
BRAF mutation, as well, is present in almost 12\% of the metastatic colon-rectal cancer, and is associated with a worse prognosis, together with an important reduction of the median survival, due to its reduced response to the therapy~\cite{RAD10,RAD15}. Finally, also the PIK3CA mutation seems to predict a resistance to EGFR antibodies therapy~\cite{RAD10}.

The testing of biomarkers needs samples of tumor tissue, by biopsy or surgery resection, and a difficultly standardizable tissue processing by pathologists that can take several weeks~\cite{Van_Cutsem_2014}. 
Moreover, the analysis of a tumor's mutational pattern can be affected by the intra-lesional heterogeneity and it may change during the systemic therapy: the archived DNA can be no more representative of the lesions sampled~\cite{Tabernero_2015,Yang_2018}. 
An alternative solution is the analysis of circulating DNA, a non-invasive and real-time technique that permits the monitoring of the changes of the mutational status~\cite{Tabernero_2015,Calandri_2020}, but it is applicable only on released DNA in circulating blood, giving only partial information~\cite{Yang_2018}. 
The development of a non-invasive technique, that allows the real-time analysis of the whole tumor, repeatable more times during systemic treatment period, could hugely improve the tailored management of oligometastatic disease.

In this work, we take into account a pioneering task, by posing baselines for predicting CRC gene mutation from only CT imaging on radial liver lesions. To our knowledge, no other DeepLearning (DL) based  studies have been proposed so far beside a single attempt by CT radiomic features modeling~\cite{Yang_2018}. 
However, Yang~\emph{et al.}~\cite{Yang_2018} method just build binary classifier by predicting if a lesion is mutated by comparing it with non-mutated samples, by aggregating KRAS/NRAS/BRAF patients and not considering PIK3CA. In Yang~\emph{et al.}, classification performances between different mutation samples is out of scope: in practice, the contribute has to be interpreted as a binary classifier for spotting non-mutated lesions. Therefore the two attempts are not strictly comparable because we explore classification between mutation classes. 
Our secondary contribution is to advance a solid pipeline to explore lesion morphology in 2D CT scan, by considering state-of-the-art efficient network architecture, attention mechanisms and semi-supervised techniques advantages. Furthermore, we hope to put in the spotlight an undervalued and challenging problem in computer-aided diagnosis.

\section{Related Works}

Starting from 2016, Garc{\'\i}a-Figueiras~\emph{et al.}~\cite{Garcia-Figueiras2016} already explores a variety of  recent developments in imaging technologies and their key role in colorectal cancer (CRC). The introduction of imaging techniques may improve the assessment of diagnosis, planning therapy, and treatment response of the patients. In particular, he reports how colorectal CT and MRI can be considered the best radiological diagnostic tests for screening CRC and polyps~\cite{unitopatho}, besides lung or liver metastases during radiological staging of CRC.  
Radiomic relevance and scanning imaging techniques are also explored in depth by Garc{\'\i}a-Figueiras~\emph{et al.}~\cite{Garcia-Figueiras2018} with a special focus on the impact of advanced liver imaging for therapy planning in CRC. More recently, Renzulli~\emph{et al.}~\cite{Renzulli2020-jv} focused his review on CRC liver metastases, considering both MRI and CT, by comparing the acquisition methods in different CRC phases.

Approaching medical image analysis, Deep Learning (DL) algorithms have become the first main tool for a multitude of tasks, from classification to semantic segmentation. Debelee~\emph{et al.}~\cite{Debelee2020-sh} reports how well DL state-of-the-art techniques can deal with tumor recognition (breast, cervical, brain, colon and lung cancers) by using different imaging methods. Debelee~\emph{et al.} shows a variety of approaches like network training from scratch and transfer learning.
An example of liver lesion and computer-aided diagnosis with Convolutional Neural Network (CNN) approaches is reported by Yamakawa~\emph{et al.}~\cite{Yamakawa2019-tx}. In his work, a classification of multiple types of carcinoma is predicted by automatically analyzing B-mode images and associated Region of Interest (ROI).
A more specific effort, in lesion detection, through CT scans is done by Yan~\emph{et al.}~\cite{Yan2018-qd}. He proposes \emph{DeepLesion}, a framework that is capable of universally detecting lesions in multiple organs, like breast and liver, and tissues. In \emph{DeepLesion}, a 3D image input is used, thanks to the huge number of patients available for the study. 
Later on, Tang~\emph{et al.}~\cite{YoubaoTang2019} improves the efficacy of \emph{DeepLesion} by proposing \emph{ULDor}, a Residual Convolutional Neural Network (R-CNN) based pipeline for general lesion detection with region proposal.
Research in lesion detection trough CT scans has become more accessible recently with the publication of annotated large datasets, e.g. as proposed by Chaudhry~\emph{et al.}~\cite{unitochest}.
In 2020, also Li~\emph{et al.}~\cite{Li2020-lo} proposes a hybrid DL pipeline specific to CRC metastasis detection from CT. In his study, he takes in consideration breast lesions only.

From the point of view of DL techniques, and specifically computer vision and image analysis, several works start to propose techniques to increase the attention of CNN models on specific features. As an example, Eppel~\emph{et al.}~\cite{SagiEppel2017} explored a supervised approach  by encoding images with their ROI annotation. Of course, unsupervised approaches to increase the network attention on hidden features have acquired a greater resonance in the field. The main reference for state-of-the-art NN attention layers, widely used in NN transformers, was  proposed in 2017 by Vaswani~\emph{et al.}~\cite{VaswaniSPUJGKP17} originally for natural language processing tasks. His contribution consists in defining a specific attention layer that helps the neural network to focus itself on relevant internal activation maps features. Attention mechanisms find application in the medical context due to the high dimension of the images generated by the acquisition processes. Tao~\emph{et al.}~\cite{Tao2019-xo} proposes a spatial attention mechanism in the processing of 3D CT scans to improve carcinoma lesion detection.

In recent years the DL landscape was dominated by a large variety of different CNN. The exploration and research in NN model architectures increases the capability of DL approaches in every aspect of computer vision. As an example, EfficentNet, a compound-scaling model proposed by Tan~\emph{et al.}~\cite{Efficientnet2019} , is one of the state-of-the-art architectures that can be used as performing backbone for general purpose image processing.
Image analysis in healthcare took a big advantage and enhancements by using more and more sophisticated NN architectures in DL based solutions.
A performance comparison between multiple residual network architecture, dealing with CT images and small datasets, is initially performed by Bagchi~\emph{et al.}~\cite{Bagchi2020}. Then Yang~\emph{et al.}~\cite{Yang2021-ge} extends the coverage of the analysis on small datasets of lung CT for COVID-19 diagnosis. He clearly shows that EfficientNet can outperform other pretrained architectures in terms of overall performances and number of parameters. In this work, we assume that EfficientNet is a valid backbone architecture for our purpose, and it will be the reference for our entire work.  

Abbas~\emph{et al.}~\cite{DeTrac2020} shows the effectiveness of transfer learning approaches in the medical image domain by using a hybrid pipeline of supervised and unsupervised techniques. With transfer learning, they were able to use a pretrained NN in a similar context, and use its set of parameters as a starting point for image extracting features. Recently, the outstanding work by Azizi~\emph{et al.}~\cite{REMEDIS} proposes a combined pipeline of self-supervised and supervised approaches, reporting performance gain using an out-of-distribution (OOD) or unannotated dataset in a similar medical domain context in order to pretrain a neural network model. It measures the efficiency in using the large quantity of unlabelled data available in medical institutes, in order to train a neural network to be able to extract valuable features from images that share the same acquisition type of the main task we want to solve. 
An efficient feature extraction algorithm, called \emph{SimCLR}, proposed by Chen~\emph{et al.}~\cite{SimClr}, is suitable for OOD datasets by using contrastive learning based technique. Contrastive learning is a self-supervised method in which a backbone is trained to recognize the same image across multiple image transformations, by uniforming the extracted features. In this work we take advantage of OOD pretrain to deal with our dataset, taking advantages of other liver lesions CT scan datasets like IRCADb~\cite{IRCADb} and LiTS17~\cite{LiTS17}.

\section{Dataset}

The data we managed to use is collected, by retroactive research, from the Picture Archiving and Communication System (PACS) of Molinette Hospital, in Turin. We will refer to our collection of colorectal carcinoma with the name \emph{CRC4AI}\footnote{Public release planned in 2023}. The dataset is collected retrospectively to five years, by including patients with advanced CRC disease. The data collection, that includes CT images, biopsies referrals and annotations, lasted about 2 months by specialized radiologists. A number of 50 patients, with a total of 92 liver lesions, is selected from the system, by following strict criteria:

\begin{itemize}

\item	Colorectal carcinoma has to be expanded and its metastases need to have  involved the liver of the patient. The carcinoma has to be visible in CT scans.
\item CT scan has to be available and the metastases need to be fully visible. The scan distance between slices needs to be at maximum 2.5 millimeters.% minimum 1 millimeter.
\item The tissue of hepatic lesions must be collected by resection or biopsy of the liver.
%The patient must have undergone a biopsy of the tumor liver tissues. 
%\francesca{i pazienti che ho segmentato io avevano fatto l'analisi genomica direttamente sul pezzo operatorio, forse sarebbe più corretto mettere una frase generale, }.
\item The patient needs to be correlated with the response of the tissue analysis by the pathological anatomy department.
\item The analysis response must be obtained within 90 days from the CT scan acquisition. It is important to take into account this constraint because the tumor can grow and change fast over the time. Therefore more mutations can appear or the lesion itself can change the morphology underlying its mutation.

\end{itemize}

\subsection{Targets and Annotations}

The pathological anatomy response contains a list of mutations involving the tissue. Depending on the genes and nucleotide positions analyzed, some results could be generated by the presence of one of several possible mutations, giving rise to multiple interpretations. In our scenario, responses include the following carcinoma mutations tests results:
\begin{itemize}
    \item K-RAS: This determination is crucial, because KRAS protein status can influence response to treatments aimed at inactivating the epidermal growth factor receptor (EGFR).
    \item N-RAS: Different RAS family mutation, with similar  consequences to K-RAS in treatments responses.
    \item PIK3CA: Mutation of a gene that acts as major regulator of several  cellular functions, such as cell proliferation, apoptosis and growth.
    \item BRAF: Protein that regulates the signals involved in cell growth. A  BRAF mutation predisposes lesions to unstable proliferation.
    \item OTHER: the tissue is negative for all the above specific tests for each considered mutation.
\end{itemize}

%\begin{table}[!ht]
%    \centering
%    \caption{Number of patients and lesions with a certain mutation. The number if images is calculated after the preprocessing step (Sec.~\ref{sec:preproc}).}
%    \label{tab:numbers}
%    \begin{tabular}{|l|c|c|c|c|c|}
%    \hline
%        ~ & N-RAS & K-RAS & BRAF & PIK3CA & OTHER \\ \hline
%        n. of patients & 2 & 21 & 3 & 9 & 22 \\ \hline
%        n. of lesions & 5 & 40 & 3 & 20 & 37 \\ \hline
%        n. of images & 110 & 575 & 55 & 228 & 728 \\ \hline
%    \end{tabular}
%\end{table}

\begin{table}[htbp]
    \centering
    \caption{Percentage of patients and lesions with a certain mutation. The number of images is calculated after the preprocessing step (Sec.~\ref{sec:preproc}).}
    \label{tab:numbers}
    \begin{tabular}{|l|c|c|c|c|c|}
    \hline
        ~ & N-RAS & K-RAS & BRAF & PIK3CA & OTHER \\ \hline
        \% of patients & 4 & 37 & 5 & 16 & 38 \\ \hline
        \% of lesions & 5 & 38 & 3 & 19 & 35 \\ \hline
        \% of images & 6 & 34 & 3 & 14 & 43 \\ \hline
    \end{tabular}
\end{table}

\begin{figure}[htbp]
    \centering
    \includegraphics[width=\columnwidth]{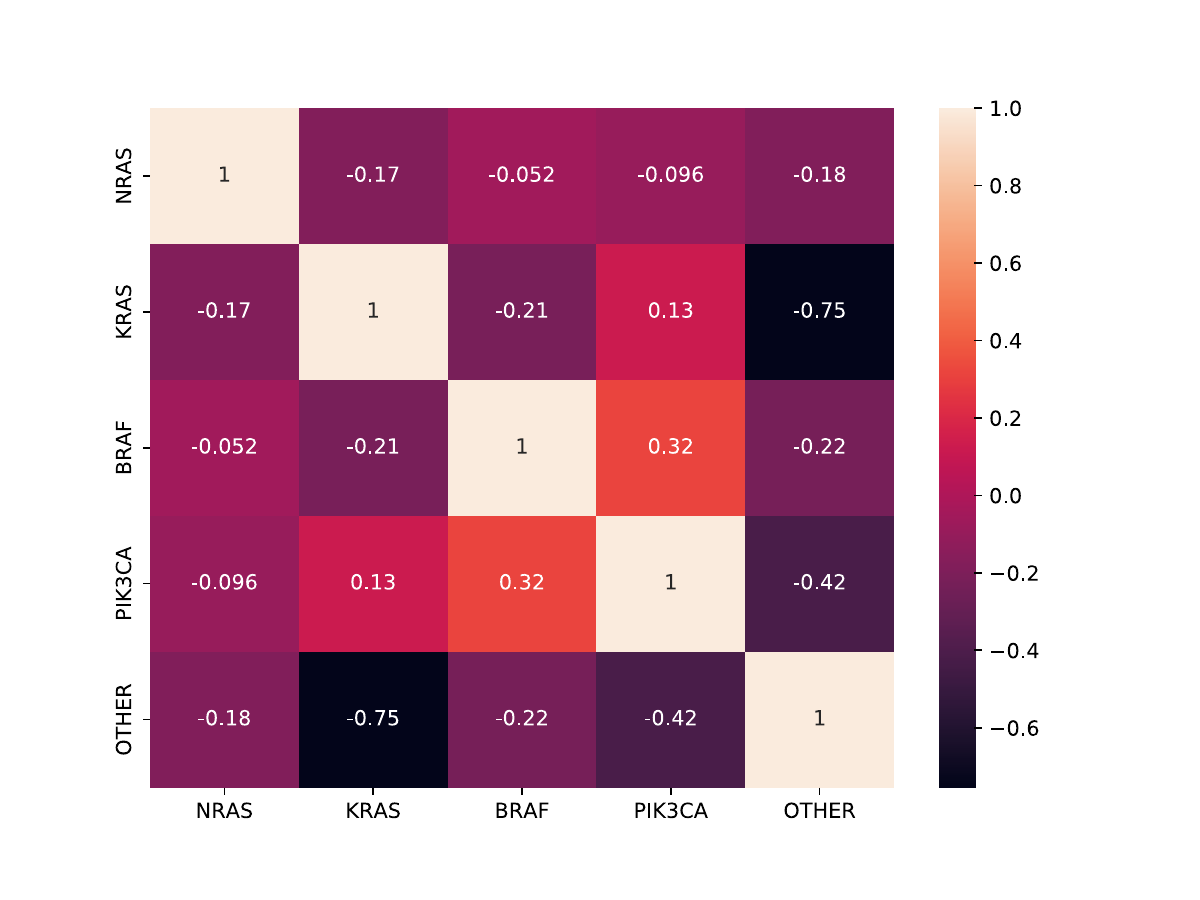}
    \caption{Correlation matrix between lesion mutations. A significant correlation can be find between PiK3CA and BRAF samples.}
    \label{fig:correlation}
\end{figure}

Some lesions may present more than one mutation at the same time.
The metastasis distribution along each mutation type is summarised in Tab.~\ref{tab:numbers}, while Fig.~\ref{fig:correlation} shows the correlation between the lesions. 

In order to address this task, our team, composed by 3 expert radiologists, annotated the entire dataset. Due to the fact that we need to look at lesion surroundings, we need to highlight the cancer lesions in each CT scan. Therefore, for each patient, the radiologists provided us all the liver lesions segmentation. Defining the CT depth $d$ as the number of slides in the exam, the annotation is a 3D binary mask (for a total of $512\times512\times d$ voxels). Each voxel of the annotation mask has a positive value if it overlaps with a lesion, 0 otherwise.

\begin{figure}[!ht]
\centering
%\captionsetup[subfigure]{labelformat=empty}
  \begin{subfigure}[b]{0.3\columnwidth}
    \centering
    \includegraphics[width=1.\linewidth]{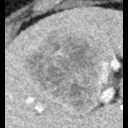}
    %\caption{Age}
    %\label{fig:91pred}
  \end{subfigure}
  %\vspace{-2\baselineskip}
  %\hfill
  %\hspace{1em}%
  \begin{subfigure}[b]{0.3\columnwidth}
    \centering
    \includegraphics[width=1.\linewidth]{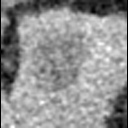}
    %\caption{Nodule Diameter}
    %\label{fig:154pred}
  \end{subfigure}
  %\hfill
  \begin{subfigure}[b]{0.3\columnwidth}
    \centering
    \includegraphics[width=1.\linewidth]{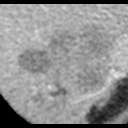}
    %\caption{Age}
    %\label{fig:1812pred}
  \end{subfigure}
  \vskip 2pt
  %\hfill
  %\hspace{1em}%
  \begin{subfigure}[b]{0.3\columnwidth}
    \centering
    \includegraphics[width=1.\linewidth]{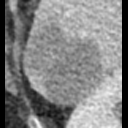}
    %\caption{Age}
    %\label{fig:45pred}
  \end{subfigure}
  %\hfill
  %\hfill
  %\hfill
  \begin{subfigure}[b]{0.3\columnwidth}
    \centering
    \captionsetup{justification=centering}
    \includegraphics[width=1.\linewidth]{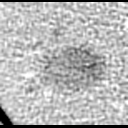}
    %\caption{Diam.=2-5mm\\IoU=0.53}
    %\label{fig:91gt}
  \end{subfigure}
  %\vskip 2pt
  %\vspace{-2\baselineskip}
  %\hfill
  \begin{subfigure}[b]{0.3\columnwidth}
    \centering
    \captionsetup{justification=centering}
    \includegraphics[width=1.\linewidth]{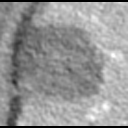}
    %\caption{Diam.=7.8mm\\IoU=0.87}
    %\label{fig:154gt}
  \end{subfigure}
  \vskip 2pt
  %\hfill
  \begin{subfigure}[b]{0.3\columnwidth}
    \centering
    \captionsetup{justification=centering}
    \includegraphics[width=1.\linewidth]{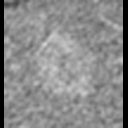}
    %\caption{Diam.=37mm\\IoU=0.93}
    %\label{fig:1812gt}
  \end{subfigure}
  %\hfill
  %\begin{subfigure}[b]{0.18\columnwidth}
  %  \centering
  %  \captionsetup{justification=centering}
  %  \includegraphics[width=1.\linewidth]{images/D1_8.png}
    %\caption{Diam.=161mm\\IoU=0.80}
    %\label{fig:45gt}
  %\end{subfigure}
\begin{subfigure}[b]{0.3\columnwidth}
    \centering
    \captionsetup{justification=centering}
    \includegraphics[width=1.\linewidth]{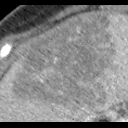}
    %\caption{a)}
    %\label{fig:45gt}
  \end{subfigure}
\begin{subfigure}[b]{0.3\columnwidth}
    \centering
    \captionsetup{justification=centering}
    \includegraphics[width=1.\linewidth]{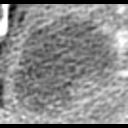}
    %\caption{b)}
    %\label{fig:45gt}
  \end{subfigure}
  \caption{Samples from data distribution: multiple examples of CRC4AI lesions, the last two images in the bottom row are taken form IRCADb and LiTS17 respectively.}
  \label{fig:samples}
\end{figure}

Due to the limited size in the number of patients and in the amount of lesions in \emph{CRC4AI}, a 3D approach to the problem is infeasible. Considering 3D lesion information as a single input, intra-patient biases can alter the results and show performances that do not correspond to an effective capacity for generalization.
Therefore, our approach is based on 2D images prediction, that is feasible with our fist version of \emph{CRC4AI}. Some examples are reported in Fig.~\ref{fig:samples}.
\section{Method}

\subsection{CT Preprocessing}
\label{sec:preproc}

CT raw data come in different formats from Molinette Hospital PACS, due to the variety of machinery used in the clinical everyday workflow%: DICOM or NIFTI.
. From the archive system, we had to retrieve 3D data, sometimes stored in 2D slices as DICOM images, sometimes stored as Neuroimaging Informatics Technology Initiative (NIFTI) file format.

The amount of total lesions in the first version of \emph{CRC4AI} is not enough to ensure significant samples variety by using their 3D information as input for a DL-based model. Therefore, we focus on 2D slices of the lesions, instead of their 3D volume, as efficient data augmentation strategy. 

Working with the entire CT scan is not useful for the task we want to solve: indeed the image contains a lot of information regarding all tissues, bones and organs of the patient, but we want to focus our analysis on the lesion only. 

\begin{algorithm}[htpb]
\caption{CT Preprocessing algorithm with 3D annotations}\label{alg:preprocessing}
\begin{algorithmic}
\Require {$\mathcal P$ pool of patients 3D CT exams,  $\mathcal A$ pool of patients 3D annotation, $\epsilon$ as exclusion slide threshold, $b$ as lesion border thickness, $r$ as the 2D resolution of the output images}
\ForEach {$p \in \mathcal P $}
\State $\mathcal A_l \gets ConnectedComponents(\mathcal A(p) )$

\Comment{$\mathcal A_l$ is the pool of different lesions annotations for $p$}
\ForEach {$a_l \in \mathcal A_l $} 
\State $\mu_l \gets mean(a_l)$

\ForEach {$a_s \in a_l $}  

\Comment{$a_s$ is the 2D slide annotation from 3D $a_l$}
\State $a_s \gets ErosionAndDilation(a_s)$ 
\If {$sliceInclusion(a_s,\mu_l,\epsilon)$}
\State $a_s \gets AddBorderThickness(a_s,b)$ 

\Comment{writing the CT crop at $a_s$ coordinates}
\State $write(p,a_s,r)$ %
\EndIf
\EndFor
\EndFor
\EndFor
\end{algorithmic}
\end{algorithm}

Our mutation predictor has to be a per-lesion classifier, so we need to take in consideration only the metastases and its surrounding tissue. Defining $b$ as the lesion border thickness, our radiologists suggest to consider $b=1$cm for surroundings inclusion~\cite{Wang_2012}. CT scans metadata contain distance information (in mm) between pixels, so a precise conversion is possible. Each lesion image is retrieved by converting the raw CT scans from Hounsfield units to grayscale range $[0,255]$ by using a window width of 400 HU (Hounsfield Units) and center of 40 HU.

In order to obtain per-lesion information from the 3D annotation map, a 3D connected components algorithm is used with 26-connected neighborhood.
It is not rare that the annotations, which are  generated with commercial 3D viewer tools, can show artifacts or gaps in their 3D volume. Therefore, in order to mitigate the problem, morphological operations such as erosion and dilation are applied on each 2D slice of the annotation.

Our radiologists, and radiomic best practices, suggest as to drop the extremes of the lesion for our analysis: the smallest regions of the lesion are often very poorly informative about the metastasis nature and do not have enough surface to infer some morphological features~\cite{Wang_2012}. Therefore, all the 2D lesion slices with area below a threshold are discarded. If we denote $l$ as the considered lesion, $a^s_l$ the annotation area in a certain slice $s$, $\mu_l$ as the mean per-slice area of the lesion, the inclusion criteria is shown below:
\begin{equation}
    sliceInclusion(s) =  a^s_l > \epsilon * \mu_l
\end{equation}
We empirically find the threshold $\epsilon =0.4$ , that ensures a good tradeoff between the image quality and the number of 2D images for each lesion. 

Alg.~\ref{alg:preprocessing} summarises the full scan preprocessing pipeline.

\begin{comment}
\begin{algorithm}[htpb]
\caption{CT Preprocessing algorithm with 3D annotations}\label{alg:preprocessing}
\begin{algorithmic}
\REQUIRE {$\mathcal P$ pool of patients 3D CT exams,  $\mathcal A$ pool of patients 3D annotation, $\epsilon$ as exclusion slide threshold, $b$ as lesion border thickness, $r$ as the 2D resolution of the output images}
\FORALL {$p \in \mathcal P $}
\STATE $\mathcal A_l \gets ConnectedComponents(\mathcal A(p) )$

\COMMENT{$\mathcal A_l$ is the pool of different lesions annotations for $p$}
\FORALL {$a_l \in \mathcal A_l $} 
\STATE $\mu_l \gets mean(a_l)$

\COMMENT{$a_s$ is the 2D slide annotation from 3D $a_l$}
\FORALL {$a_s \in a_l $}  
\STATE $a_s \gets ErosionAndDilation(a_s)$ 
\IF {$sliceInclusion(a_s,\mu_l,\epsilon)$}
\STATE $a_s \gets AddBorderThickness(a_s,b)$ 

\COMMENT{writing the CT crop at $a_s$ coordinates}
\STATE $write(p,a_s,r)$ %
\ENDIF
\ENDFOR
\ENDFOR
\ENDFOR
\end{algorithmic}
\end{algorithm}
\end{comment}

\subsection{Self-Supervised Pretraining}
% Transfer learning is a widely used technique in machine learning, especially in computer vision, which consist in using a pretrained network for a different task. Typically, a network trained for image classification is used for tasks involving medical images. 

% To go beyond this standard approach to transfer learning, we used SimCLR \luca{citazione}, which is a framework for self-supervised pretraining.

Transfer learning is a widely used technique in machine learning, especially in computer vision, which allows to use a pretrained network on a different task. This method is extremely advantageous if the available dataset for training is limited and allows to reduce the time needed for training, since only finetuning of the weights is necessary.

\begin{table}[htbp]
\begin{center}
\caption{Liver lesions datasets composition}
\label{table:datasets}
\begin{tabular}{c @{\extracolsep{\fill}} ccccc}
\hline\noalign{\smallskip}
 & Name & Patients & Lesions & Total Images\\
\noalign{\smallskip}
\hline
\noalign{\smallskip}
& CRC4AI & 50 & 92 & 1063\\
& IRCADb & 15 & 118 & 817 \\
& LiTS17 & 130 & 885 & 10466\\
\hline
\end{tabular}
\end{center}
\end{table}

In our case, we used SimCLR~\cite{SimClr}, which is a framework for self-supervised contrastive learning, in order to pretrain a network using images from similar datasets. We chose two publicly available datasets of liver cancer CT images: IRCADb~\cite{IRCADb} and LiTS17~\cite{LiTS17,LiTS17ext}, originally designed for segmentation tasks. We refer to overall external data as Out-Of-Distribution (OOD) dataset. Our aim is to improve the performance of the classifier by using a network that was pretrained on CT liver images, therefore similar to the ones in our CRC4AI dataset, on top of a more traditional one pretrained with natural images~\cite{REMEDIS}. 
For this purpose, SimCLR is very fitting, since self-supervised pretraining does not require labeled images and no additional work on the public datasets was needed. 
SimCLR works as follows:
\begin{itemize}
    \item Firstly, we preprocess IRCADb and LiTS17 with the same algorithm used for CRC4AI. IRCADb contains 15 useful patients with 118 annotated lesions while LiTS17 is a larger dataset that includes 130 patients and 885 annotated samples. A distribution recap is shown in Tab.~\ref{table:datasets}. Samples from both datasets are joined together (OOD dataset) to compose our training SimCLR image pool.
    \item Secondly, stochastic augmentation of the images is performed by applying random cropping with resizing, random flipping, random color distortion and random Gaussian blur.
    \item Thirdly, a neural network base encoder extracts representation vectors from the augmented data. In our case, we used EfficientNet-B0~\cite{Efficientnet2019} for this purpose. The model will be used as pretrained backbone for CRC mutation classification.
    \item Lastly, a multilayer perceptron (MPL) with one hidden layer is used as a projection head to map the representation vectors to the space where a contrastive loss function can be applied to maximize the agreement between the differently augmented views of the same image~\cite{SimClr}.
\end{itemize}
After this pretraining, the base encoder is used as the backbone of the final network to be used for our task. Since longer trainings are observed to be beneficial to the backbone capability to extract relevant features~\cite{SimClr}, we opted to use a training cycle with 1000 epochs.

% This procedure was repeated a second time adding a self-attention layer at the end of the base encoder before the MPL of SimCLR in order to verify whether this addition can improve the classification performance. After the pretraining, the new base encoder made up of EfficientNet-B0 plus the self-attention layer is used as the backbone of the final network. 

\subsection{Genomics mutation classification}
\label{sec:classification}

We perform a multi-label classification starting from the 2D scans, by using the backbone pretrained in the self-supervised step. Due to the nature of the diagnostic response, that we used as our ground-truth labels, multiple mutations can be present in a single lesion. 

A binary cross-entropy loss is used to minimize the error on the model output probabilities. Each data mini-batch passes trough data multiple augmentation transformations, like random rotation, image flipping, image cropping and then tensor normalization. 
\ \\
\ \\
There are two important considerations in order to have a fair performance measurement:
\begin{itemize}
 \item We want to avoid intra-patient biases between validation and training set, in order to prevent the model to learn specific patient features for measuring predictions quality. Therefore, samples from a target patient can only be present in the same set: the intersection of the patients between test and train set is strictly empty. 
 \item We want to avoid inter-patient variability to be a bias in our measurements. Some patients have many more different lesions than others: straightforward by dealing with a small dataset, a single patient can have a significant impact on the model training or on the result measurements.
\end{itemize}
In order to avoid these possible issues, we opted for a multi-seeded K-Fold validation method: each random seed perform a samples split (90\% training and 10\% test) by maintaining the patient data separation. Furthermore, we also ensure the labels distribution consistency in the process. The results are then measured by averaging the performances obtained from each initialization.

Due to the natural imbalance in data distribution typical for medical datasets, not mitigated from the K-Fold split strategy, we adopt a simple data-balancing strategy.
At the beginning of the training phase, train-set images are re-sampled or sub-sampled in relation to their mutation. We use the label distribution median in order to equalize the number of samples for each mutation.
\ \\

\section{Results}

In order to highlight the complexity of the task, we will report multiple strategies as comparison:
\begin{itemize}
\item a baseline, by using EfficientNet as encoder, pratrained on ImageNet without self-supervised pretrain.
The last fully connected classification layer is substituted by a new one that fits the desired output dimensions. 
\item a second model (OODp), by using the \emph{SimCLR}-pretrained model on OOD samples. The MLP head is removed and replaced. The final architecture is the same as the first baseline.

\item a last model with a slight architecture change, by including a self attention (SA) mechanism in order to evaluate its impact on lesion images. We repeat both experiments ahead, by including a multihead attention layer, followed by ReLU activation function, before the classification fully connected layer. In the contrastive pretraining scenario, the attention layer is inserted before the model MLP head proper of SimCL approach. 
\end{itemize}

\begin{figure}[!ht]
\centering
\begin{subfigure}[b]{.8\columnwidth}
    \centering
    \includegraphics[width=\linewidth]{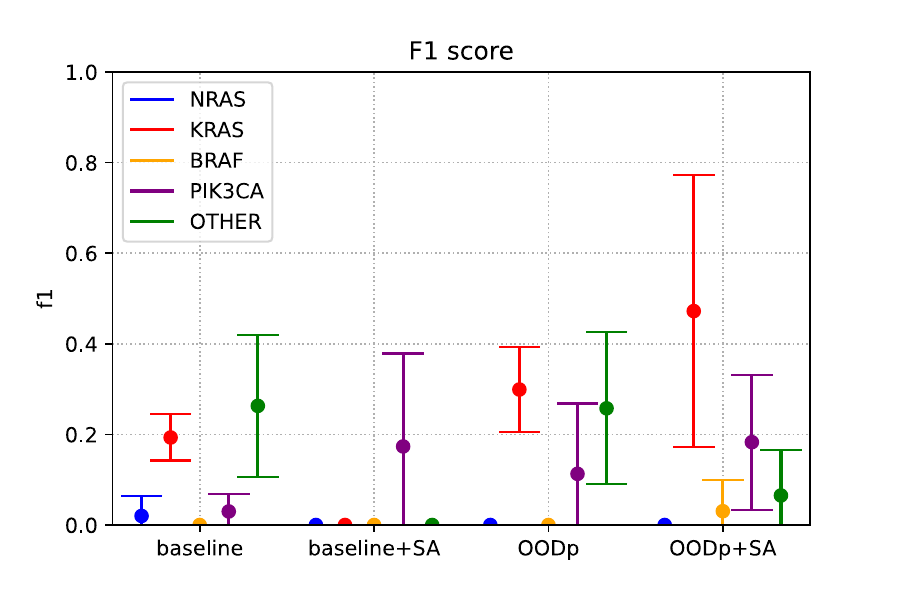}
    \caption{F1 Score, 5 classes }
    \label{fig:5f1_128}
\end{subfigure}
\begin{subfigure}[b]{.8\columnwidth}
    \centering
    \includegraphics[width=\linewidth]{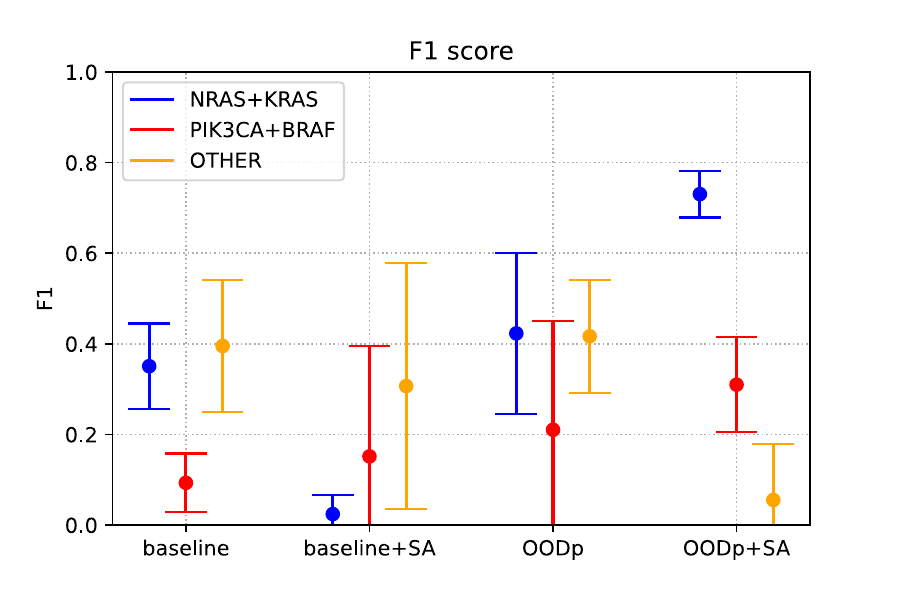}
    \caption{F1 Score, 3 classes}
    \label{fig:3f1_128}
\end{subfigure}
\begin{subfigure}[b]{.8\columnwidth}
    \centering
    \includegraphics[width=\linewidth]{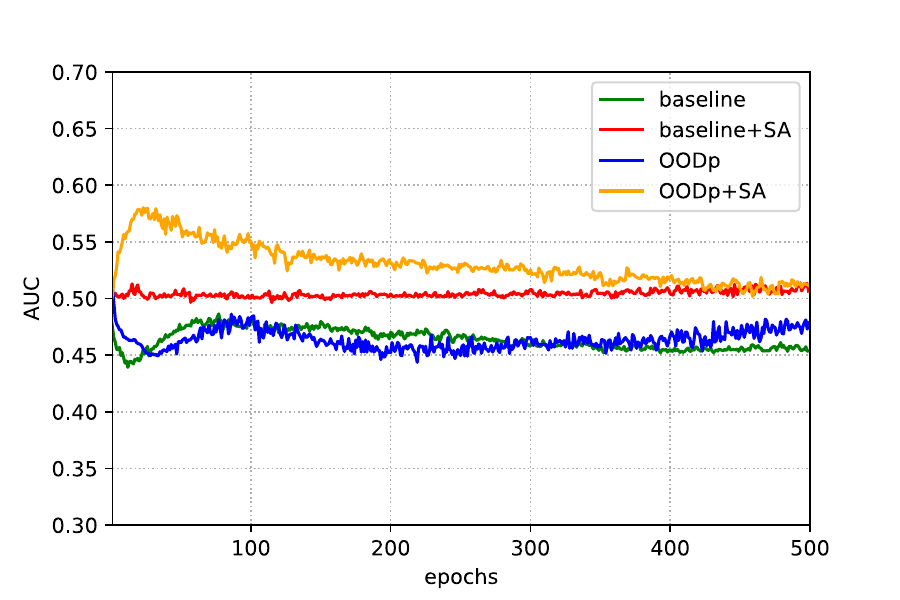}
    \caption{AUC, 5 classes}
    \label{fig:5auc_128}
\end{subfigure}
\begin{subfigure}[b]{.8\columnwidth}
    \centering
    \includegraphics[width=\linewidth]{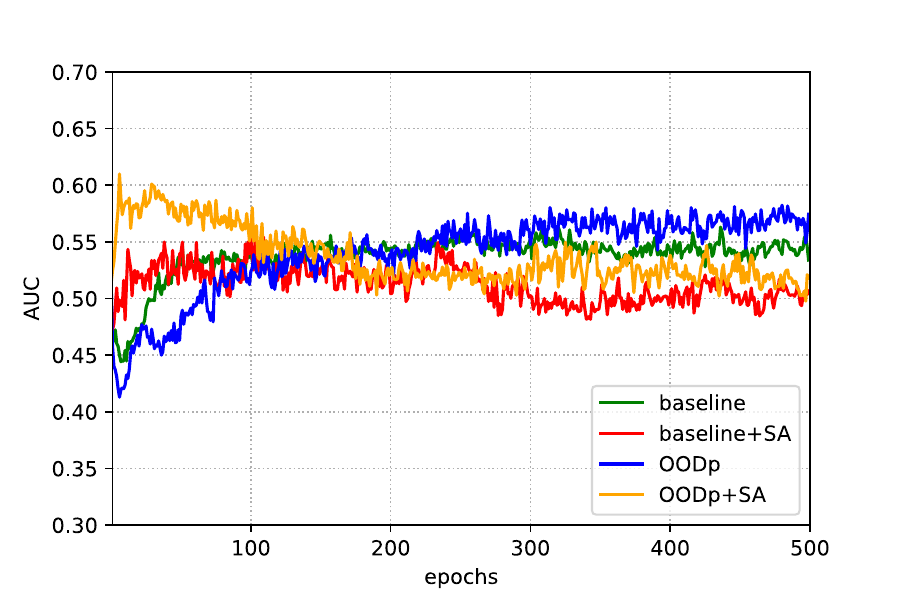}
    \caption{AUC, 3 classes}
    \label{fig:3auc_128}
\end{subfigure}
\caption{Comparison of F1 score and AUC between the classifications with 5 and 3 classes. We can notice how OODp+SA is more prone to overfit without any early-stopping policy.}
\label{fig:f1-auc}
\end{figure}

All the experiments are executed by using 2 GPUs NVIDIA GeForce RTX 2080 SUPER, and share the same hyper parameters and the same \emph{PyTorch} implementation. The network training hyper parameters are empirically retrieved by grid search. We found the optimal settings, equal for each test, using $0.001$ as learning rate, weight decay \num{5e-4}, a mini-batch size of 16 samples and SGD as optimizer, with $0.9$ momentum. All the experiments are repeated and averaged with 5 different seeds for data splits, as reported in section~\ref{sec:classification}. 
%Furthermore, we use multiple input resolutions ($32 \times 32$, $64 \times 64$, $128 \times 128$ pixels) in order to evaluate possible impacts on performances.
\ \\
\ \\
In Fig.~\ref{fig:5f1_128} and Fig.~\ref{fig:3f1_128} we can see the F1 scores retrieved for the 5 classes task. 
All the results are reported with a confidence level of $95\%$ ($p$ value $< 0.05$), metrics are derived by considering 5 different initialization  seeds.
It is straightforward to see the difficulty of the task to discriminate all the mutations at once using such a small dataset. Therefore, we opted for composing new categories by grouping the mutations in 3 classes: NRAS+KRAS, PIK3CA+BRAF and OTHER.
N-RAS and K-RAS mutations share the same RAS mutation root, while PIK3CA and BRAF show the highest correlation between lesions.
In this way, we also partially compensated the CRC4AI imbalance, as can be seen in Tab.~\ref{tab:numbers}, of the different classes and we obtained an improvement of AUC score, especially for the baseline and the OODp models (comparison between Fig.~\ref{fig:5auc_128} and Fig.~\ref{fig:3auc_128} scores). The AUC score is derived by using "One-vs-rest" policy. We compute the AUC of each class against the rest, by including weighted averaging for class imbalance.

\begin{figure}[htpb]
\centering
\begin{subfigure}[b]{0.75\columnwidth}
    \includegraphics[width=\linewidth]{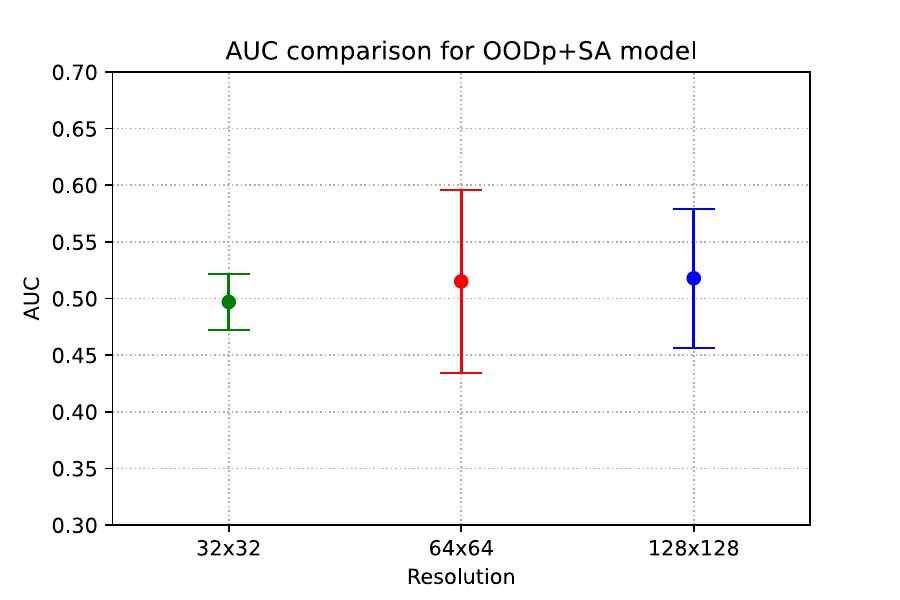}
    % \caption{AUC, resolution 32$\times$32}
\end{subfigure}
\begin{subfigure}[b]{0.75\columnwidth}
   \includegraphics[width=\linewidth]{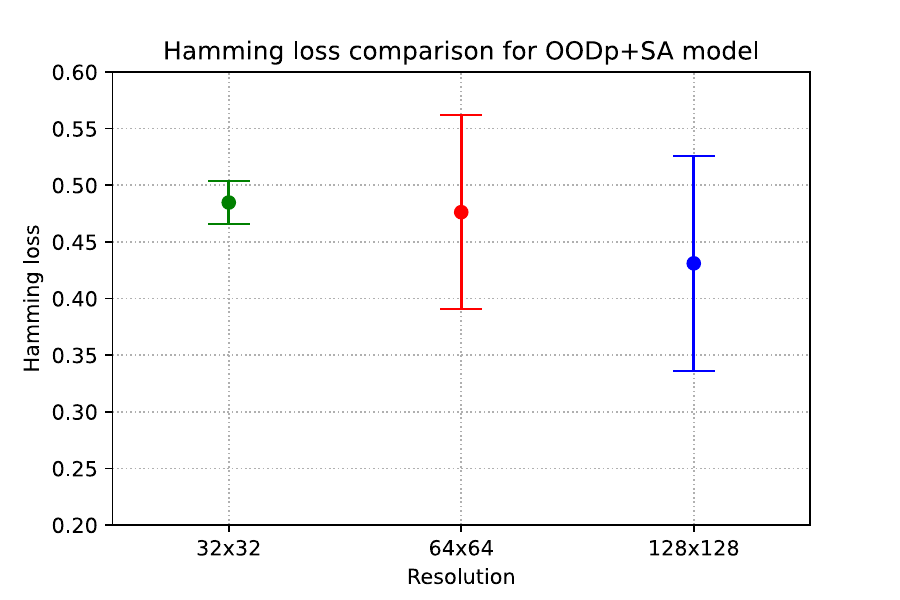}
%   \caption{AUC, resolution 64$\times$64}
\end{subfigure}
\caption{Comparison of AUC and Hamming loss(lower is better) of the OODp+SA model with the different resolutions. The small variance between different input resolutions indicates a model preference for coarse morphological features, rather than finer image details.}
\label{fig:auc_hamming}
\end{figure}

\begin{figure}[htpb]
    \centering
    \begin{subfigure}[b]{0.72\columnwidth}
        \includegraphics[width=\linewidth]{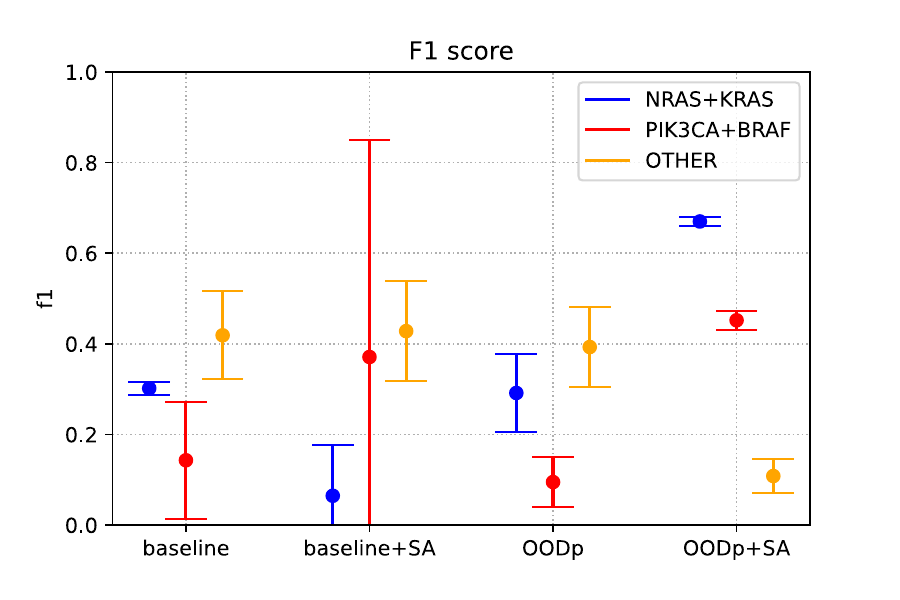}
        \caption{F1 score, resolution 32$\times$32}
    \end{subfigure}
    \begin{subfigure}[b]{0.72\columnwidth}
        \includegraphics[width=\linewidth]{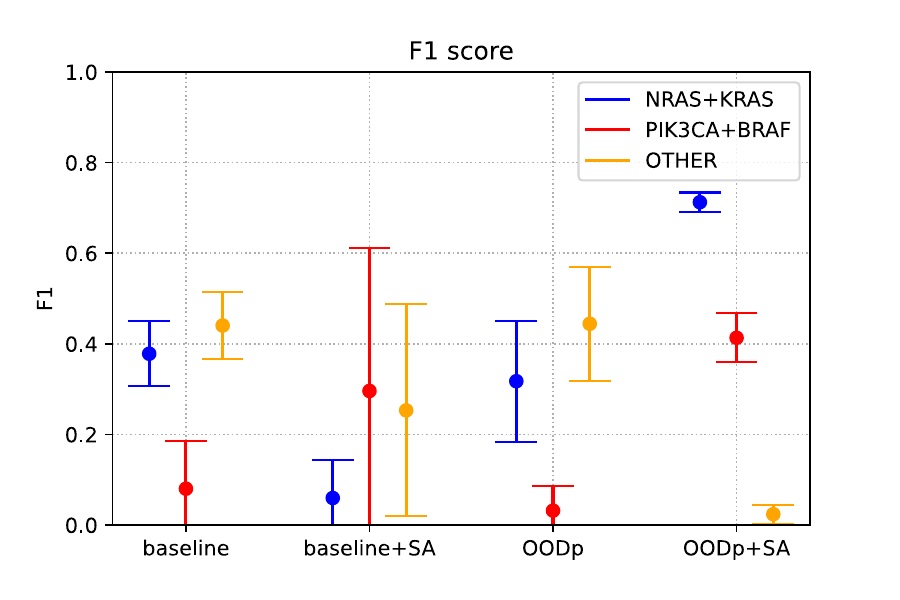}
        \caption{F1 score, resolution 64$\times$64}
    \end{subfigure}
    \begin{subfigure}[b]{0.72\columnwidth}
        \includegraphics[width=\linewidth]{images/f1_128x128_svg-tex.pdf}
        \caption{F1 score, resolution 128$\times$128}
    \end{subfigure}
    \caption{Comparison among the F1 scores of the different models with different image resolutions. In each scenario, a promising RAS classification capability is evident for OODp+SA model.}
    \label{fig:f1_res}
\end{figure}

We then evaluated the effect of image resolution on the classification capability of the different models. In order to do so, we sampled the images of the cropped lesions at three different resolutions: 32$\times$32, 64$\times$64 and 128$\times$128 pixels. As can be seen in Fig.~\ref{fig:auc_hamming} and Fig.~\ref{fig:f1_res}, the difference between the 32$\times$32 and 64$\times$64 resolutions is not significant, but the  highest one (128$\times$128) seems to slightly improve the results, especially concerning AUC, F1 score and Hamming loss. Tab.~\ref{tab:auc_f1_res}, reports the values in detail for each model. 
Since there is a certain variability in the size of the metastases, our hypothesis is that the information content of an image of a small lesion may not be affected by the oversampling at a high resolution, whereas a larger lesion may lose information if it is subsampled. Our assumption is that coarse morphological features have slightly more relevance for CRC genomics classification rather than finer image details. Anyway, a higher resolution may be recommended for such a task, even though the difference in the performance of the classificator does not change drastically.

From the results, we can say that the pretraining with OOD datasets can improve the classification capability, compared with the baseline model. Especially with the highest resolution, the OODp and OODp+SA models perform better than, respectively, the baseline and baseline+SA model. Moreover, as can be seen in Table \ref{table:RAS}, the pretrained models have more balanced values of specificity and sensitivity. 

The most significant result of this study, however, is the F1 score of the RAS class obtained with the OODp+SA model, as can be seen in Fig.~\ref{fig:3f1_128} and Table \ref{tab:auc_f1_res}:
with a resolution of 128$\times$128 pixels we achieved $F1=0.73\pm0.05$. This means that, even with a very limited dataset, the model can recognize at least this mutation from the others. 

\begin{table*}[htpb]
    \centering
    \caption{Comparison of AUC and F1 score of the RAS isolation obtained with the different models at different resolutions}
    \begin{tabular}{|l|c|c|c|c|c|c|}
    \hline
        ~ & \multicolumn{2}{|c|}{32$\times$32}  & \multicolumn{2}{|c|}{64$\times$64} & \multicolumn{2}{|c|}{128$\times$128} \\ \hline
        ~ & AUC & F1 RAS & AUC & F1 RAS & AUC & F1 RAS  \\ \hline
        baseline & 0.49$\pm$0.04 & 0.30$\pm$0.01 & 0.52$\pm$0.05 & 0.38$\pm$0.07 & 0.53$\pm$0.05 & 0.35$\pm$0.09  \\ \hline
        baseline+SA & 0.53$\pm$0.09 & 0.06$\pm$0.11 & 0.52$\pm$0.05 & 0.06$\pm$0.08 & 0.51$\pm$0.01 & 0.02$\pm$0.04  \\ \hline
        OODp & 0.49$\pm$0.04 & 0.29$\pm$0.09 & 0.50$\pm$0.06 & 0.32$\pm$0.13 & \textbf{0.57$\pm$0.07} & 0.42$\pm$0.18  \\ \hline
        OODp+SA & 0.50$\pm$0.03 & 0.67$\pm$0.01 & 0.52$\pm$0.08 & 0.71$\pm$0.02 & \textbf{0.52$\pm$0.06} & \textbf{0.73$\pm$0.05}  \\ \hline
    \end{tabular}
    \label{tab:auc_f1_res}
\end{table*}

\begin{table*}[h]
    \centering
    \caption{In depth classification performances for RAS mutation family (N-K RAS). The combination on Out Of Distibution pretrain plus Self Attention layer can reach the best RAS Specificity/Sensitivity compromise at higher input resolutions.}
    \label{table:RAS}
    \begin{tabular}{|l|c|c|c|c|c|c|}
    \hline
        ~ & \multicolumn{2}{|c|}{32$\times$32}  & \multicolumn{2}{|c|}{64$\times$64} & \multicolumn{2}{|c|}{128$\times$128} \\
        \hline
        ~ & Specificity & Sensitivity & Specificity & Sensitivity & Specificity & Sensitivity  \\ \hline
        baseline & 0.74$\pm$0.21 & 0.21$\pm$0.04 & 0.78$\pm$0.20 & 0.27$\pm$0.07 & 0.88$\pm$0.12 & 0.23$\pm$0.07  \\ \hline
        baseline+SA & 0.97$\pm$0.04 & 0.04$\pm$0.06 & 0.96$\pm$0.07 & 0.03$\pm$0.05 & 1.00$\pm$0.00 & 0.01$\pm$0.02  \\ \hline
        OODp & 0.81$\pm$0.1 & 0.20$\pm$0.06 & 0.76$\pm$0.10 & 0.23$\pm$0.11 & 0.89$\pm$0.03 & 0.30$\pm$0.14  \\ \hline
        OODp+SA & 0.08$\pm$0.01 & 0.88$\pm$0.04 & 0.21$\pm$0.20 & 0.90$\pm$0.09 & \textbf{0.34$\pm$0.08} & \textbf{0.83$\pm$0.16}  \\ \hline
    \end{tabular}
\end{table*}

\section{Conclusion}

In this work, we present our attempt of solving CRC mutation classification from CT images of liver metastasises. We explore multiple state-of-the art solutions for approaching this extremely challenging problem. We know how important it is to establish the colon tumour mutation as soon as possible for patients follow-up and therapeutic strategies. As highlighted above, it is essential to remember that this type of classification, and diagnosis by expert doctors, is not possible without biopsy on patients. Consequently this work stands as a basis for novel non-invasive solutions in the context of genomics identification of CRC metastases.

With our approach, we try to answer the question by taking into account medical imaging data. It is straightforward that, because of our really small dataset, this is a preliminary work and further investigations on all the CRC mutations have to be considered. We are convinced that collecting more data is the way to enable this study to take into consideration 3D convolution models and improve overall performances. Furthermore, an enlarged dataset can be the key to classify other rare genomics mutations beside the more common RAS mutation.

Despite the low number of patients, obtaining an image-based RAS classifier with 0.73 F1 score is more than we expected from our research in raw CT inspections. Considering the promising results, our radiologists are currently starting to evaluate the possibility to include a classifier like the one we developed in their daily workflow, though being aware that better results will be needed before a systematic clinical application. An AI-based classifier could allow to start to apply therapeutic strategies while waiting the necessary days for the response from the liver tissue biopsy analysis. 

\newpage
\bibliographystyle{splncs04}
\bibliography{egbib}
\end{document}